\documentclass[runningheads]{llncs}
\usepackage{caption}
\usepackage{subfigure}
\usepackage[T1]{fontenc}

\usepackage{cite}
\usepackage{amsmath,amssymb,amsfonts}
\usepackage{algorithmic}
\usepackage{graphicx}
\usepackage{textcomp}
\usepackage{xcolor}
\usepackage{color, colortbl}
\definecolor{Gray}{gray}{0.9}
\usepackage{amsmath}
\usepackage{pifont}
\newcommand{\cmark}{\ding{51}}%
\newcommand{\xmark}{\ding{55}}
\usepackage{makecell}
\usepackage{wrapfig}
\usepackage{picinpar}

\usepackage[utf8]{inputenc} %
\usepackage[T1]{fontenc}    %
\usepackage{hyperref}       %
\usepackage{url}            %
\usepackage{booktabs}       %
\usepackage{amsfonts}       %
\usepackage{nicefrac}       %
\usepackage{microtype}      %

\usepackage{algorithmic}
\usepackage{graphicx}
\usepackage{textcomp}
\usepackage{xcolor}
\usepackage{multirow}
\usepackage{algorithm}
\usepackage{bm}
\usepackage{subfigure}
\usepackage{caption}

\usepackage{CJKutf8}

\usepackage{graphicx}
\begin{document}
\title{Prompt-based Effective Input Reformulation\\for Legal Case Retrieval}

\author{Yanran Tang \and Ruihong Qiu \and Xue Li}
\authorrunning{Y. Tang et al.}
\institute{The University of Queensland\\
\email{\{yanran.tang, r.qiu\}@uq.edu.au, xueli@eecs.uq.edu.au}}

\maketitle              %
\begin{abstract}
Legal case retrieval plays an important role for legal practitioners to effectively retrieve relevant cases given a query case. Most existing neural legal case retrieval models directly encode the whole legal text of a case to generate a case representation, which is then utilised to conduct a nearest neighbour search for retrieval. Although these straightforward methods have achieved improvement over conventional statistical methods in retrieval accuracy, two significant challenges are identified in this paper: (1) Legal feature alignment: the usage of the whole case text as the input will generally incorporate redundant and noisy information because, from the legal perspective, the determining factor of relevant cases is the alignment of key legal features instead of whole text matching; (2) Legal context preservation: furthermore, since the existing text encoding models usually have an input length limit shorter than the case, the whole case text needs to be truncated or divided into paragraphs, which leads to the loss of the global context of legal information. In this paper, a novel legal case retrieval framework, PromptCase, is proposed to tackle these challenges. Firstly, \textit{\textbf{legal facts}} and \textit{\textbf{legal issues}} are identified and formally defined as the key features facilitating legal case retrieval based on a thorough study of the definition of relevant cases from a legal perspective. Secondly, with the determining legal features, a prompt-based encoding scheme is designed to conduct an effective encoding with language models. Extensive zero-shot experiments have been conducted on two benchmark datasets in legal case retrieval, which demonstrate the superior retrieval effectiveness of the proposed PromptCase. The code has been released on \url{https://github.com/yanran-tang/PromptCase}.

\keywords{Legal case retrieval \and Information retrieval}
\end{abstract}

\section{Introduction}
Legal case retrieval (LCR) aims to retrieve relevant cases given a query case, which is important for legal practitioners in the world's two major legal systems, common law and civil law. From a legal perspective, the precedents are the historical cases that are similar to a given case in two determining aspects,  legal facts and legal issues. In common law system, the judicial reasons of a judgement are critically based on the relevant cases, which is also called ``the doctrine of precedents''~\cite{precedent}. While in civil law system, although the judgement is not necessarily to be based on previously relevant cases, judges and lawyers are still strongly suggested to obtain legal information from these relevant cases~\footnote{\url{https://www.court.gov.cn/zixun-xiangqing-243981.html}}. Nowadays, the methods of LCR can be generally divided into two branches, statistical retrieval models~\cite{BM25,TF-IDF,LMIR} that measure the term frequency similarity between cases and neural LCR models~\cite{Law2Vec,Lawformer,MTFT-BERT,MVCL,BERT-PLI,LEGAL-BERT,SAILER,JOTR,DoSSIER,RPRS, NOWJ,ConversationalAgent,UA@COLIEE2022,CL4LJP} that encode the case into a representation to conduct nearest neighbour search. 

Recently, neural LCR models have greatly attracted the research focus for the outstanding text representation ability. Generally, BERT-based legal case retrieval models use the whole text of a case to obtain the representation of the case, which directly determines the retrieval quality and accuracy with the case similarity calculation afterwards~\cite{Law2Vec,Lawformer,MTFT-BERT,MVCL,LEGAL-BERT}. Due to the input length limit of BERT-based models, e.g., 512 tokens~\cite{BERT}, a case is typically too long to be directly fed into these models with more than 5,000 words in a case. Therefore, most of these methods rely on truncating the case text to a suitable length~\cite{SAILER} or dividing the whole text into smaller segments to process the input text~\cite{BERT-PLI}.

Although these models have achieved competitive progress compared to traditional statistical models in LCR, there are still two challenges remaining: (1) Legal feature alignment: using the whole case text as the input for case representation generation~\cite{Law2Vec,Lawformer} will incorporate redundant and noisy information, because from legal perspective, the determining factor of relevant cases is the alignment of legal features instead of whole text matching. When legal practitioners are retrieving relevant cases, they are actually finding ``precedents'', which refer to a court decision in an earlier case with legal facts and legal issues similar to the current case~\footnote{\url{https://www.uscourts.gov/glossary}}. Therefore, similar legal facts and legal issues are the key to retrieving relevant cases considering legal theory. (2) Legal context preservation: furthermore, the whole case text is usually truncated~\cite{SAILER} or divided into paragraphs~\cite{BERT-PLI} due to the input length limit of BERT-based models, which is ineffective in capturing the legal context information. A legal case generally contains more than 5,000 words (in certain situations, easily exceeding 50,000 words), which is much longer than the 512-token input limit for BERT~\cite{BERT}, 16k-token for Longformer~\cite{longformer}, or 8k-token for ChatGPT~\cite{chatgpt}. Thus, passively truncating or dividing the case 
will lead to a significant loss of decisive legal features and case global view among the legal context information.

In light of the above observations, a novel LCR framework called PromptCase is proposed in this paper to tackle these challenges. Firstly, the input representation with two determining legal features, legal facts and legal issues, are proposed to effectively obtain representative legal information in cases instead of using the entire case. According to the formal legal document writing requirements, the format of a case text is well structured so that \textit{\textbf{legal facts}} and \textit{\textbf{legal issues}} can be effectively extracted from the case with sufficient processing steps. Secondly, in order to effectively encode the extracted legal features, a novel prompt-based encoding scheme is proposed to encode these features with language models. Empirical experiments are conducted on two benchmark datasets, LeCaRD~\cite{LeCaRD} and COLIEE~\cite{COLIEE}, which shows that the specific legal features proposed in this paper can represent the legal case more precisely to make a good representation for neural LCR models and effectively improve the performance of neural LCR models. The main contributions of this paper are summarised as follows:
\begin{itemize}
\item A PromptCase model is proposed for effective legal case retrieval by tackling the legal feature alignment and legal context preservation challenges.
\item Two determining legal features, \textit{\textbf{legal facts}} and \textit{\textbf{legal issues}} are identified and extracted from legal cases with adequate processing procedures.
\item A prompt-based encoding scheme is derived to effectively encode the extracted legal features for the widely used language models.
\item Extensive experiments conducted on two benchmark datasets demonstrate the state-of-the-art performance of the PromptCase framework.
\end{itemize}

\section{Related Work}
\subsection{Legal Case Retrieval}
LCR is a special type of IR. The methods of IR can be generally divided into two branches, statistical methods~\cite{TF-IDF,BM25,LMIR} and neural network methods~\cite{ColBERT,Doc2query,Sentence-BERT,crossencoder}. Similarly, in LCR there are the same two branches. Statistical models include TF-IDF~\cite{TF-IDF}, BM25~\cite{BM25} and LMIR~\cite{LMIR}, which rely on the term frequency and inverse document frequency of words to determine the similarity between cases. Neural LCR models rely on encoding the case using the language models~\cite{DeepCT,BERT,RoBERTa,monot5}. With the increasing amount of online legal information and users’ legal information needs, many neural LCR models~\cite{Law2Vec,Lawformer,MTFT-BERT,MVCL,BERT-PLI,LEGAL-BERT,SAILER,JOTR,DoSSIER,RPRS,NOWJ,ConversationalAgent,UA@COLIEE2022,IOT-Match,Law-Match} are conducted to bridge the information gap by capturing domain-specific and personal needs. Law2Vec~\cite{Law2Vec} is a legal language model that pre-trains on a large legal corpus. Lawformer~\cite{Lawformer} focuses on combining three types of attention mechanisms to get the context of long legal cases. BERT-PLI~\cite{BERT-PLI} calculates the similarity between two paragraphs of cases text to tackle the lengthy problem of legal cases. SAILER~\cite{SAILER} is a pre-trained language model that selects the reasoning, decision and fact sections in the cases to train the encoder and uses the fact section to be the input of the encoder to get the case representation. 

\subsection{Input Reformulation in Neural Legal Case Retrieval}
Input reformulation plays an important role in neural LCR because a case is hard to fit into the model directly due to the length limit~\cite{SAILER,LEDsummary,BM25injtct,LeiBi,LEVEN,JNLP@COLIEE2019}. Askari et al.~\cite{LEDsummary} and LeiBi~\cite{LeiBi} both propose to combine lexical and neural network methods to get a summary of a legal case as the case representation. LEVEN~\cite{LEVEN} utilises the frequency of legal events to reformulate the case input. Both CL4LJP~\cite{CL4LJP} and QAjudge~\cite{QAjudge} intuitively reformulate the case input with only the fact instead of the whole case. IOT-Match~\cite{IOT-Match} reformulates the case input based on legal rationales. BERT-PLI~\cite{BERT-PLI} divides the case input into the paragraph-level interaction between query and candidate cases. Liu et al.~\cite{action_conversational_agent,query_conversational_agent} proposes to use the conversational search paradigm to reformulate the query case.

\section{Preliminary}

\subsection{Task Definition}
\label{sec: task}
In legal case retrieval, given a query case $q$, and a set of $n$ candidate cases, denoted as $\mathcal{D}=\{d_1,d_2,...,d_n\}$, the task is to retrieve a set of relevant cases $\mathcal{D}^* = \{d^*_i| d^*_i \in \mathcal{D} \wedge relevant (d^*_i, q) \}$ from $\mathcal{D}$, where $relevant (d^*_i, q)$ denotes that $d^*_i$ is a relevant case of the query case $q$. From a legal perspective, the relevant cases are called precedents, which are the historical cases with legal facts and legal issues similar to the given query case. Specifically, given a query case, the relevant cases in COLIEE2023 dataset are the cases referred by the query case. While in LeCaRD dataset, cases having similar key facts and key circumstances to the query case are labelled as relevant cases by legal experts.

\subsection{Input Reformulation in Neural Legal Case Retrieval}
\label{sec: case_representation}
Existing neural LCR models generally use the full case as the input to the model with different input reformulation methods to deal with the overly long cases.

\paragraph{BERT-PLI}~\cite{BERT-PLI} reformulates the case input into the paragraph-level interaction vector between the query and candidate cases as below: 
\begin{equation} \label{eq:BERT-PLI_1}
    \mathbf{e}_{(q_i,d_j)} = \text{BERT}([\text{CLS}] ; q_i ; [\text{SEP}] ; d_j ; [\text{SEP}]), 
\end{equation}
where ``$;$'' denotes the concatenation function, and [CLS] and [SEP] are two special tokens for BERT to denote the input's beginning and separation. $q_i$ and $d_j$ are the $i$-th paragraph and  $j$-th paragraph of case $q$ and $d$. 

\paragraph{SAILER}~\cite{SAILER} uses the fact section of cases as the input of a finetuned BERT:
\begin{equation}
\label{eq:SAILER}
    \mathbf{e}_q = \text{BERT}([\text{CLS}];q_{(\text{fact})}; [\text{SEP}]),\quad\mathbf{e}_d = \text{BERT}([\text{CLS}];d_{(\text{fact})}; [\text{SEP}]),        
\end{equation}
where $q_{(\text{fact})}$ and $d_{(\text{fact})}$ are the fact of the query case and the document case respectively. The fact is assumed to be located in the most front and if the length of the fact is longer than 512 tokens, the first 512 tokens of the case will be used.

\paragraph{BM25Inject}~\cite{BM25injtct} concatenates the BM25 score of the query case and the document case into the input of BERT-based cross-encoder:
\begin{equation} \label{eq:BM25injtct}
    s_{(q,d)} = \text{BERT}([\text{CLS}] ; q ; [\text{SEP}] ; s_{\text{BM25}_{(q,d)}} ; [\text{SEP}] ; d ; [\text{SEP}]),
\end{equation}
where $s_{\text{BM25}_{(q,d)}}$ is the BM25 score scalar of the query case $q$ and the candidate $d$ and the final semantic similarity is $s_{(q,d)} $.

\section{Method}
\label{sec: methods}
\begin{figure}[!t]
\centering
\includegraphics[width=\textwidth]{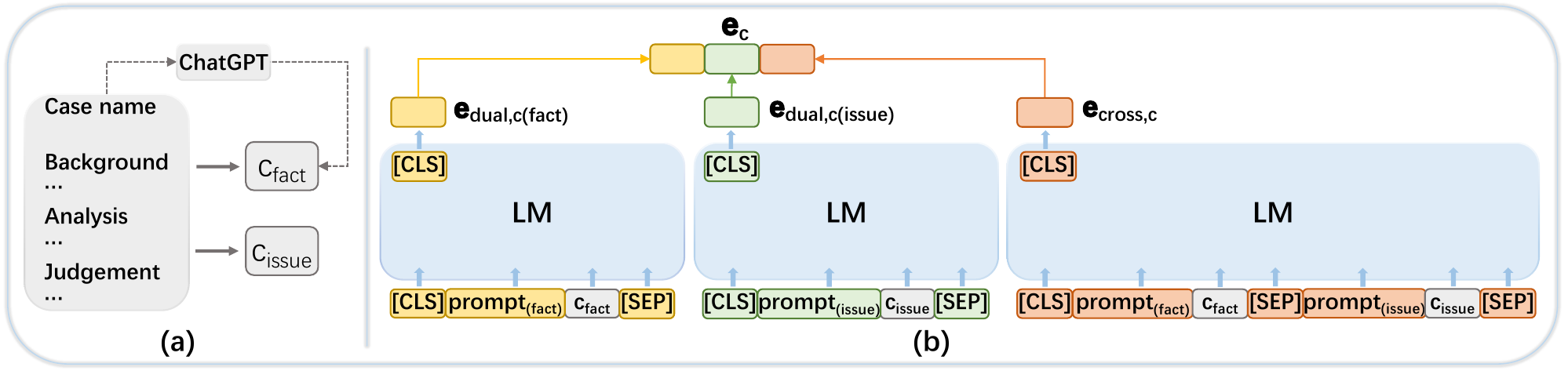}
\caption{The framework of PromptCase. LM means a language model, e.g., BERT. The final output of the [CLS] token is the representation embedding of a legal fact, a legal issue or a case. (a) The process of legal facts and legal issues extraction. When legal facts are not explicitly available, ChatGPT is applied to generate a case summary as legal facts. (b) Dual and cross encoding with prompt of a case.} 
\label{fig:PromptCase}
\end{figure}

In this section, the PromptCase framework will be introduced. In Section~\ref{sec: features}, two determining legal features are extracted. A prompt-based method utilising these legal features will be detailed in Section~\ref{sec: representation}. The measurement of case similarity will be introduced in Section~\ref{sec: similarity}. The overview of PromptCase is shown in Fig.~\ref{fig:PromptCase}.

\subsection{Extraction of Legal Facts and Legal Issues}
\label{sec: features}
\begin{figure}[!t]
\centering
    \subfigure[COLIEE dataset (common law)]{
    \includegraphics[width=0.45\linewidth]{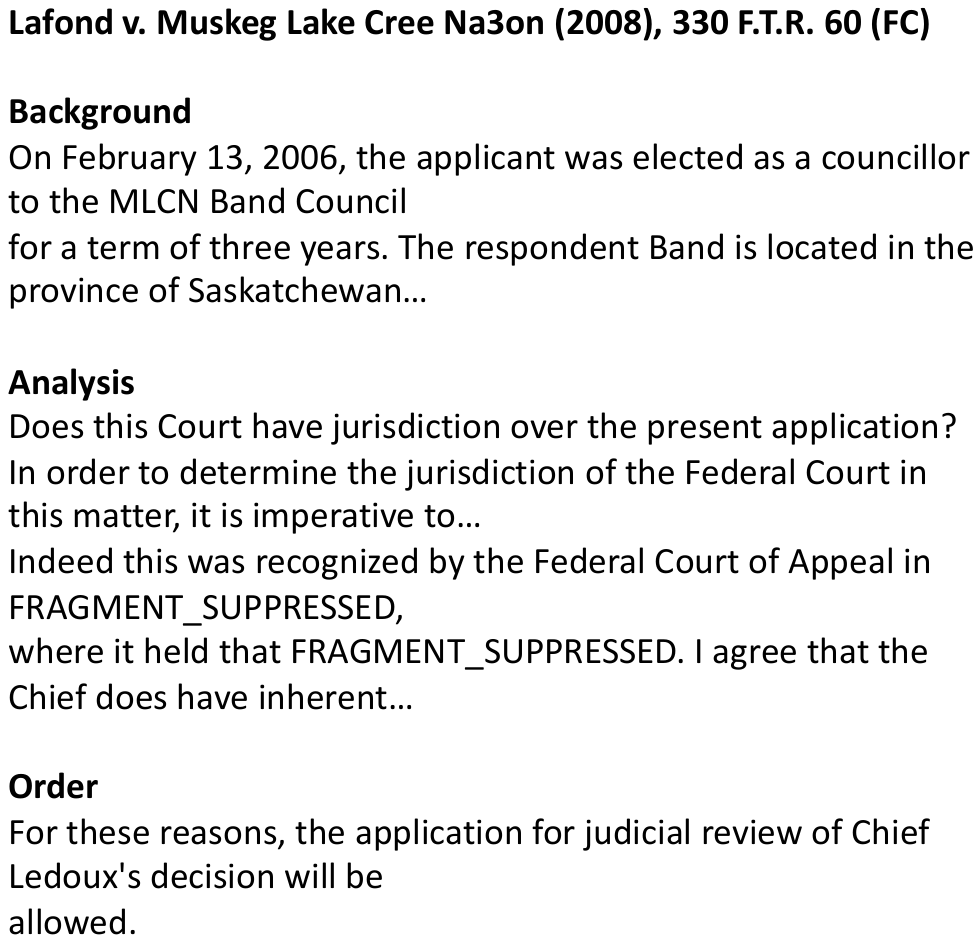}
    \label{fig:coliee_example}
    }
    \subfigure[LeCaRD dataset (civil law)]{
    \includegraphics[width=0.45\linewidth]{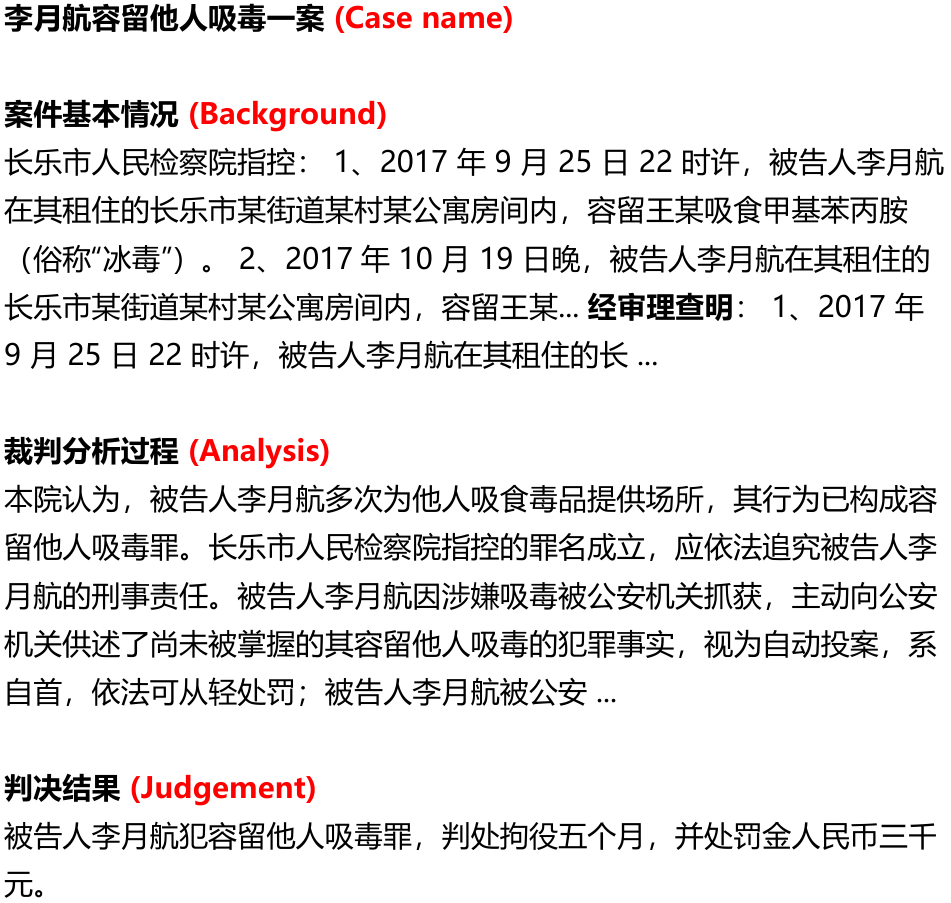}
    \label{fig:lecard_example}
    }
\caption{Example of case documents}
\label{fig:case_example}
\end{figure}

This section describes the extraction of legal facts and legal issues from cases as shown in Fig.~\ref{fig:PromptCase}(a) to overcome the legal feature alignment challenge. For common law (COLIEE dataset) or civil law (LeCaRD dataset) respectively, a case often has a relatively fixed writing style, which includes four basic parts as in Fig.~\ref{fig:case_example}. The first part is the case name with basic information about the case. The second part is the ``Background'' of the case demonstrating detailed information about the case. The third part is ``Analysis'' describing the reasons why the judges make the final decision. The final part called ``Order'' or ``Judgement'', is the judgement of the case. Such a clear and general structure of legal cases provides access to locate and extract legal facts and legal issues from extremely long cases.
\vspace{-5pt}

\subsubsection{Legal facts.}
\label{sec: facts}
Legal fact is a fundamental part that describes the ``who, when, what, where and why'' in legal cases. Firstly, in the COLIEE2023 dataset, the detailed process of a case is generally written in the background part, which is often more than thousands of words that will exceed the input limit of BERT-based models. In order to get an abstract yet accurate legal facts of cases, the ChatGPT~\cite{chatgpt} is used to get the summary of legal facts. The ChatGPT API with ``gpt-3.5-turbo'' model is used with the prompt of ``Summarise in 50 words:~''. As a result, the output of ChatGPT will be the legal facts $c_{(\text{fact})}$ of the case $c$. 

\begin{CJK*}{UTF8}{gbsn}
Secondly, in LeCaRD, the fact section is a separate and brief part that can be found in ``Background'', beginning with a description of ``After the trial, it was found out that: '' in Chinese (the bold Chinese words ``\textbf{经审理查明}：'' in the ``Background'' part in Fig.~\ref{fig:lecard_example}). Thus, in LeCaRD, the legal facts $c_{(\text{fact})}$ of the case $c$ are extracted directly based on the understanding of a legal case.
\end{CJK*}

\subsubsection{Legal issues.}
\label{sec: issues}
The definition of ``issue'' in legal domain is ``a critical feature that focuses on the dispute points between the parties in the case.''\footnote{\url{https://www.uscourts.gov/glossary}} In case documents of common law, the legal issues are located in the ``Analysis'' part, which is given by the judges to settle the disputes between the parties with legal reasons. To have convincing reasons, the judges will list the relevant precedents' facts, issues or judgements in this part to support the judges' opinions. Specifically, as shown in Fig.~\ref{fig:coliee_example}, there are words replaced by placeholders with special terms in cases of the COLIEE2023 dataset, such as ``FRAGMENT\_SUPPRESSED''. The original words for these placeholders are the case name of a precedent. These placeholders are for the task of legal case retrieval, which is to find the precedents being referred in the placeholder. Thus, for the COLIEE dataset, all of the sentences with placeholders will be selected as the legal issues $c_{(\text{issue})}$ of the case $c$. 

Compared to common law, the judges in the civil law system often make their judgements according to the legal articles written in the acts while the judges of the common law system have the compulsory responsibility to refer the precedents to support their final decisions. And there is also no specific part for settling legal issues in the cases of civil law. After a thorough study of the cases from LeCaRD dataset under the civil law system, it is found that legal issues often appear in the case as the name of charges, such as ``murder''. Therefore, the names of charges in Chinese criminal law are collected and saved as a list of charges. For every case (queries and candidates) in LeCaRD, the full text of a case will be used to find the charges that appear both in the case and the list of charges. Finally, all of the found charges are the legal issues $c_{(\text{issue})}$ of the case $c$. 

\subsection{Prompt-based Case Encoding}
\label{sec: representation}
After extracting legal facts and legal issues, a prompt-based case encoding method is developed in this section to tackle the legal context preservation challenge.

\subsubsection{Prompt template.}
\label{prompt}
\begin{CJK*}{UTF8}{gbsn}
With the recent advances of prompt, the capability of prompting a language model is impressive in understanding the context information of a task. To enable the language models to capture the global context of legal information, the prompt templates of ``Legal facts:'' (``法律事实：'' in Chinese) and ``Legal issues:'' (``法律纠纷：'' in Chinese) will be added to the beginning of the legal facts and legal issues texts and fed into the language model together. For every legal case in COLIEE2023 and LeCaRD datasets, the prompt template is formulated as below:
\begin{equation}
\label{eq:prompt_dual}
    \text{prompt}_{{(\text{fact})}}= \text{``Legal facts:''},\quad\text{prompt}_{{(\text{issue})}}= \text{``Legal issues:''} .
\end{equation} 
\end{CJK*}

\subsubsection{Dual encoding with prompt.}
\label{dual-encoder}
To avoid the undesired cross-effect between legal facts and legal issues, the legal facts with prompt and legal issues with prompt will be fed into the BERT-based encoder separately to get the individual legal facts embedding $\mathbf{e}_{\text{dual},c_{(\text{fact})}}$ and legal issues embedding $\mathbf{e}_{{\text{dual},c_{(\text{issue})}}}$. The encoding process can be denoted as the following equations:
\begin{equation} \label{eq:dual_features}
\begin{split}
    \mathbf{e}_{\text{dual},c_{(\text{fact})}} &= \text{LM}([\text{CLS}] ; \text{prompt}_{(\text{fact})} ; c_{(\text{fact})} ; [\text{SEP}]), \\
    \mathbf{e}_{\text{dual},c_{(\text{issue})}} &= \text{LM}([\text{CLS}] ; \text{prompt}_{(\text{issue})} ; c_{(\text{issue})} ; [\text{SEP}]),
\end{split}
\end{equation} 
where $\mathbf{e}_{\text{dual},c_{(\text{fact})}}$ and $\mathbf{e}_{\text{dual},c_{(\text{issue})}}$ are both the embedding of the final hidden state of the [CLS] token of the language model (LM), e.g., BERT.

\subsubsection{Cross encoding with prompt.}
\label{cross-encoder}
On the contrary, to obtain the deeper interactions between legal facts and legal issues, the cross encoding method is also being conducted as the following equations:
\begin{equation} \label{cross_features}
    \mathbf{e}_{\text{cross},c} = \text{LM}([\text{CLS}] ; \text{prompt}_{(\text{fact})} ; c_{(\text{fact})} ; [\text{SEP}] ; \text{prompt}_{(\text{issue})} ; c_{(\text{issue})} ; [\text{SEP}]).
\end{equation}
where $\mathbf{e}_{\text{cross},c}$ is the output embedding of the [CLS] token of LM.

\subsubsection{Case representation}
\label{case-representation}
To obtain both the original and interaction information of legal facts and legal issues, the case representation will be the concatenation of the $\mathbf{e}_{\text{dual},c_{(\text{fact})}}$, $\mathbf{e}_{\text{dual},c_{(\text{issue})}}$, and $\mathbf{e}_{\text{cross},c}$ as the following equations:
\begin{equation} \label{case_representation}
    \mathbf{e}_{c} = \mathbf{e}_{\text{dual},c_{(\text{fact})}}; \mathbf{e}_{\text{dual},c_{(\text{issue})}}; \mathbf{e}_{\text{cross},c}.
\end{equation}

\subsection{Case Similarity}
\label{sec: similarity}
Similar to traditional IR tasks, the dot product (denoted as ($\cdot$)) is used to measure the semantic similarity between two cases. Given the case representation $\mathbf{e}_{q}$ and $\mathbf{e}_{d}$ of case $q$ and candidate case $d$ generated by PromptCase, the similarity score $s_{(q,d)}$ is calculated as:
\begin{equation} 
\label{eq:dot_product}
    s_{(q,d)} = \mathbf{e}_{q} \cdot \mathbf{e}_{d}.   
\end{equation}

\section{Experiments}
\begin{wraptable}{r}{0.45\linewidth}\centering
\caption{Statistics of LeCaRD and COLIEE2023 datasets.}
\label{tab: dataset}
\resizebox{1\linewidth}{!}{
\begin{tabular}{c|cc}
\toprule
Datasets &LeCaRD &COLIEE2023\\\midrule
\cmidrule{2-3}
Language &Chinese &English   \\
Avg. length/case &8,275 &5,566 \\
Largest length of cases &99,163 &61,965 \\
Avg. relevant cases/query &10.33 &2.69 \\
\bottomrule
\end{tabular}
}
\end{wraptable}

\subsection{Setup}
\label{sec:setup}
\subsubsection{Datasets.}
To evaluate the proposed PromptCase, the experiments are conducted on the following LCR datasets with summarised statistics in Table~\ref{tab: dataset}.

\textbf{LeCaRD}~\cite{LeCaRD}. LeCaRD is a legal case retrieval dataset, where the cases are from the supreme court of China, a civil law system country. It contains 107 queries and over 43,000 candidate cases. For each query, there is a candidate pool of 100 cases. The evaluation of LeCaRD is based on the binary golden label for a more restrict requirement~\footnote{\url{https://github.com/myx666/LeCaRD\#golden_labelsjson}}.

\textbf{COLIEE2023}~\cite{COLIEE}~\footnote{\url{https://sites.ualberta.ca/~rabelo/COLIEE2023/}}. COLIEE2023 is a dataset from Competition on Legal Information Extraction/Entailment (COLIEE) 2023, where cases are from the federal court of Canada with common law system. Given a query case, relevant cases are retrieved from the entire candidate pool. To avoid the data leakage problem of pre-trained models, only the testing set of COLIEE2023 is used.

\subsubsection{Metrics.}
\label{metrics}
For both datasets, precision (P), recall (R), Micro F1 (Mi-F1), Macro F1 (Ma-F1), Mean Reciprocal Rank (MRR), Mean Average Precision (MAP) and normalized discounted cumulative gain (NDCG) are used. For both LeCaRD and COLIEE2023 datasets, top 5 ranking results are evaluated by following previous methods~\cite{LeCaRD,SAILER,COLIEE}. All metrics are the higher the better.

\subsubsection{Baselines.}
\label{baselines}
The following baselines are chosen for comparison:
\begin{itemize}
    \item \textbf{BM25}~\cite{BM25} is a statistical retrieval model using the term frequency and inverse document frequency, which is still a strong baseline.
    \item \textbf{BERT}~\cite{BERT} is a strong bi-directional transformer encoder in language tasks. For LeCaRD in Chinese, the ``uer/sbert-base-chinese-nli''~\cite{BERT-chinese} model is used , while for COLIEE2023 in English, the ``bert-base-uncased''~\cite{BERT} model is used.
    \item \textbf{Lawformer}~\cite{Lawformer} is pre-trained on Chinese legal corpus and focuses on long documents processing.
    \item \textbf{LEGAL-BERT}~\cite{LEGAL-BERT} is pre-trained on a large English legal corpus and achieves state-of-the-art results in different legal understanding tasks.
    \item \textbf{MonoT5}~\cite{monot5} is a pre-trained sequence-to-sequence model focuses on document ranking task using the powerful T5 model~\cite{T5}.
    \item \textbf{SAILER}~\cite{SAILER} is a structure-aware pre-trained model that achieves state-of-the-art performance on both datasets. Two-stage usage of SAILER with BM25 is evaluated as well.
\end{itemize}

BERT-PLI~\cite{BERT-PLI} is not compared since its paragraph-level interaction is not applicable to legal facts and legal issues. BM25Inject~\cite{BM25injtct} is not compared because its cross encoding between cases is not extendable in our scenario.

\subsubsection{Implementation.}
\label{implementation}
The French text in COLIEE2023 is removed. The two-stage method is based on the top 10 retrieved cases by BM25 model. All experiments are in a zero-shot manner without training, except that the SAILER model for COLIEE2023 is pre-trained on the COLIEE2023 training set. The experiment of BM25 model with PromptCase reformulated input utilises the original text, legal facts, legal issues and prompt together.

\subsection{Overall Performance}
In this section, the PromptCase is evaluated by being integrated into the baselines. The results are presented in Table~\ref{tab:LeCaRD} for LeCaRD and Table~\ref{tab:COLIEE2023} for COLIEE2023.
\begin{wraptable}{r}{0.6\linewidth}
    \centering
	\begin{minipage}{1\linewidth}
		\centering
        \vspace{-11pt}
        \caption{Overall performance on LeCaRD (\%).}\label{tab:LeCaRD}
        \resizebox{1\linewidth}{!}{
        \begin{tabular}{l|ccccccc}
        \toprule
        \multirow{2}{*}{Methods} &\multicolumn{7}{c}{LeCaRD@5} \\
        \cmidrule{2-8}
        &P@5 &R@5 &Mi-F1 &Ma-F1 &MRR@5 &MAP &NDCG@5\\\midrule
        BM25 &40.0 &19.2 &26.0 &30.5 &58.3 &48.5 &45.9 \\
        \rowcolor{Gray}
        +PromptCase &41.3 &19.9 &26.8 &31.7 &60.6 &58.8 &65.2 \\\midrule
        BERT &38.7 &18.6 &25.1 &26.7 &57.4 &54.3 &61.0 \\
        \rowcolor{Gray}
        +PromptCase &46.2 &22.2 &30.0 &35.4 &64.4 &61.2 &67.9 \\\midrule
        Lawformer &29.0 &13.9 &18.8 &19.5 &43.6 &41.9 &48.2\\
        \rowcolor{Gray}
        +PromptCase &38.9 &18.7 &25.3 &30.7 &62.0 &59.7 &64.0 \\\midrule
        SAILER &46.7 &22.5 &30.4 &37.1 &67.9 &65.4 &70.1  \\
        \rowcolor{Gray}
        +PromptCase &51.6 &24.8 &33.5 &43.0 &71.1 &67.6 &74.2 \\\midrule
        Two-stage\\
        SAILER &47.8 &23.0 &31.1 &36.1 &67.3 &64.4 &70.6 \\
        \rowcolor{Gray}
        +PromptCase &51.0 &24.6 &33.2 &38.7 &70.7 &67.9 &73.5 \\
        \bottomrule
        \end{tabular}}
    \end{minipage}
	\\ \hspace*{\fill} \\
    \begin{minipage}{1\linewidth}
        \centering
        \caption{Overall performance on COLIEE (\%).}\label{tab:COLIEE2023}
        \resizebox{1\linewidth}{!}{
        \begin{tabular}{l|ccccccc}
        \toprule
        \multirow{2}{*}{Methods} &\multicolumn{7}{c}{COLIEE2023}\\
        \cmidrule{2-8}
        &P@5 &R@5 &Mi-F1 &Ma-F1 &MRR@5 &MAP &NDCG@5  \\\midrule
        BM25 &16.5 &30.6 &21.4 &22.2 &23.1 &20.4 &23.7 \\
        \rowcolor{Gray}
        +PromptCase &17.0 &31.5 &22.1 &23.0 &24.2 &21.6 &24.4 \\\midrule
        BERT &2.07 &3.84 &2.69 &2.57 &5.51 &5.48 &6.25 \\
        \rowcolor{Gray}
        +PromptCase &2.38 &4.42 &3.10 &3.02 &6.33 &6.25 &7.21\\\midrule
        LEGAL-BERT &4.64 &8.61 &6.03 &6.03 &11.4 &11.3 &13.6 \\
        \rowcolor{Gray}
        +PromptCase &4.83 &8.96 &6.28 &6.44 &13.4 &13.4 &15.5\\\midrule
        MonoT5 &0.38 &0.70 &0.49 &0.47 &1.17 &1.33 &0.61 \\
        \rowcolor{Gray}
        +PromptCase &0.56 &1.05 &0.73 &0.72 &1.63 &1.43 &0.89\\\midrule
        SAILER &12.8 &23.7 &16.6 &17.0 &25.9 &25.3 &29.3 \\
        \rowcolor{Gray}
        +PromptCase &16.0 &29.7 &20.8 &21.5 &32.7 &32.0 &36.2\\\midrule
        Two-stage\\
        SAILER &19.6 &32.6 &24.5 &23.5 &37.3 &36.1 &40.8 \\
        \rowcolor{Gray}
        +PromptCase &21.8 &36.3 &27.2 &26.5 &39.9 &38.7 &44.0 \\
        \bottomrule
        \end{tabular}}
        \end{minipage}
\end{wraptable}
\vspace{-11pt}

Overall, the PromptCase can steadily improve the performances of all baselines by a large margin. With the state-of-the-art pre-trained SAILER model in legal domain, PromptCase significantly boosts the retrieval performance for both one and two stage manners with a proper reformulation of case input. For the traditional method BM25, the performance of using PromptCase is better than with the whole case as input. The improved performance shows that the reformulated input can capture the determining legal features with proper emphasis on the term frequency without being biased by the long and noisy case texts. For the pre-trained BERT with full case as input, the performances on both datasets are worse than BM25 and SAILER. However, BERT+PromptCase outperforms BM25+PromptCase and the SAILER baseline model on LeCaRD, which indicates that BERT is a semantic LM that can better understand and represent a case using legal features semantics. While the term frequency cannot fully take advantage of the semantics in legal facts and legal issues, which also limits the performance of two-stage SAILER on LeCaRD with or without PromptCase compared with one-stage SAILER. Lawformer and LEGAL-BERT are two neural LCR models pre-trained on Chinese and English respectively, whose performances are improved significantly with PromptCase. The performance of MonoT5 is the worst in COLIEE2023 dataset, possibly because MonoT5 is pre-trained for text-to-text tasks different from retrieval tasks. Comparing the results on these two datasets, the improvement with PromptCase on LeCaRD is more obvious than on COLIEE2023. The possible reason is the different definitions of relevance in these datasets. For LeCaRD, the relevant cases are defined by legal experts, which is easier for models to identify. While for COLIEE2023, the relevant cases are referred cases by the query case, which are a subset of all relevant cases and not a golden label for relevance, leading to an inferior performance.

\begin{table}[!t]\centering
\caption{Ablation study. Leg-Feat denotes legal features. (\%)}\label{tab:ablation}
\resizebox{1\linewidth}{!}{
\begin{tabular}{c|c|ccccccc|ccccccc}
\toprule
\multirow{2}{*}{Prompt} &\multirow{2}{*}{Leg-Feat} &\multicolumn{7}{|c|}{LeCaRD} &\multicolumn{7}{c}{COLIEE2023}\\
\cmidrule{3-16}
& &P@5 &R@5 &Mi-F1 &Ma-F1 &MRR@5 &MAP &NDCG@5 &P@5 &R@5 &Mi-F1 &Ma-F1 &MRR@5 &MAP &NDCG@5 \\
\midrule
\xmark &\xmark &46.7 &22.5 &30.4 &37.1 &67.9 &65.4 &70.1 &12.8 &23.7 &16.6 &17.0 &25.9 &25.3 &29.3 \\
\midrule
\cmark &\xmark &46.5 &22.4 &30.2 &36.9 &68.6 &65.8 &70.5 &12.8 &23.7 &16.6 &17.0 &25.4 &24.8 &28.5\\
\midrule
\xmark &\cmark &52.0 &25.0 &33.8 &43.3 &69.4 &66.2 &72.9 &15.9 &29.5 &20.6 &21.3 &32.6 &31.5 &35.8\\
\midrule
\cmark &\cmark &51.6 &24.8 &33.5 &43.0 &71.1 &67.6 &74.2 &16.0 &29.7 &20.8 &21.5 &32.7 &32.0 &36.2\\
\bottomrule
\end{tabular}}
\end{table}

\subsection{Ablation Study}
\begin{CJK*}{UTF8}{gbsn}
The ablation study is conducted to verify the effectiveness of the two main components of PromptCase, the legal features and the prompt encoding scheme. The SAILER~\cite{SAILER} model is used as the base model in these experiments since SAILER is a state-of-the-art pre-trained model with English and Chinese on both datasets. Specifically, the prompt templates of the experiment without legal features are reformulated as ``Legal facts and legal issues:'' in English and ``法律事实和法律纠纷：'' in Chinese for COLIEE2023 and LeCaRD respectively.
\end{CJK*}

As shown in Table~\ref{tab:ablation}, the reformulated input with prompt and legal features can significantly improve the performance compared with other variants for both datasets. The legal features alone can largely increase the retrieval performance. While only using prompt encoding, the performance is not improved since there is no specific legal feature used with the prompt.

\subsection{Effectiveness of Legal Features}
To verify the effectiveness of legal features, experiments are conducted using SAILER with: no legal features, only legal facts, only legal issues and both legal facts and legal issues. For no legal features, the second result in Table~\ref{tab:ablation} is reused.

\begin{table}[!t]\centering
\caption{Effectiveness of legal facts (Facts) and legal issues (Issues). (\%)}\label{tab:effect_leg_feat}
\resizebox{1\linewidth}{!}{
\begin{tabular}{c|c|ccccccc|ccccccc}
\toprule
\multirow{2}{*}{Facts} &\multirow{2}{*}{Issues} &\multicolumn{7}{|c|}{LeCaRD} &\multicolumn{7}{c}{COLIEE2023}\\
\cmidrule{3-16}
&  &P@5 &R@5 &Mi-F1 &Ma-F1 &MRR@5 &MAP &NDCG@5 &P@5 &R@5 &Mi-F1 &Ma-F1 &MRR@5 &MAP &NDCG@5 \\
\midrule
\xmark &\xmark &43.0 &20.7 &27.9 &34.0 &63.5 &51.7 &51.7 &12.8 &23.7 &16.6 &17.0 &25.4 &24.8 &28.5 \\
\midrule
\cmark &\xmark &47.3 &22.8 &30.7 &37.4 &66.8 &63.6 &70.1 &12.7 &23.5 &16.5 &17.2 &24.7 &24.3 &27.7\\
\midrule
\xmark &\cmark &41.3 &19.9 &26.8 &32.9 &57.6 &54.7 &61.7 &13.4 &24.8 &17.4 &17.8 &29.1 &28.3 &31.8\\
\midrule
\cmark &\cmark &51.6 &24.8 &33.5 &43.0 &71.1 &67.6 &74.2 &16.0 &29.7 &20.8 &21.5 &32.7 &32.0 &36.2\\
\bottomrule
\end{tabular}}
\end{table}

As shown in Table~\ref{tab:effect_leg_feat}, the reformulated input with both legal facts and legal issues achieves the best performance in the effectiveness of legal features experiments, which indicates the challenge of legal feature alignment is well resovled. For LeCaRD, the performance of only using legal facts is better than only using legal issues, while it is opposite in COLIEE2023 that only using legal issues is better than only using legal facts. This opposite phenomenon also appears in the experiments of ablation study. The different performances of datasets could be due to the different case structures in different legal systems, which may cause the different focuses of prompt and legal features.

\subsection{Effectiveness of Prompt}
In this experiment, the effectiveness of Prompt is investigated with different prompt templates using SAILER. The prompt templates are widely chosen from instructive, misleading and irrelevant categories, which are detailed in Table~\ref{tab:effect_prompts}.

\begin{table}[!t]\centering
\caption{Effectiveness of different prompts. Instructive (IT): \textbf{A}: ``Legal facts:/Legal issues:''; \textbf{B}: ``The following is legal facts:/The following is legal issues:'';\textbf{C}: ``The judge think:''; Misleading (ML): \textbf{D}: ``This case is related to \$<randomly sample one issue>:''; \textbf{E}: ``Legal facts of this case is \$<randomly sample one issue>:/Legal issues of this case is \$<randomly sample one issue>:''; Irrelevant (IR): \textbf{F}: ``Let's look:''; \textbf{G}: ``ADC is a database conference:'' and \textbf{NA}: no prompt is used.}
\label{tab:effect_prompts}
\resizebox{1\linewidth}{!}{
\begin{tabular}{c|c|ccccccc|ccccccc}
\toprule
&&\multicolumn{7}{c|}{LeCaRD} &\multicolumn{7}{c}{COLIEE2023}\\
\cmidrule{3-16}
 & &P@5 &R@5 &Mi-F1 &Ma-F1 &MRR@5 &MAP &NDCG@5 &P@5 &R@5 &Mi-F1 &Ma-F1 &MRR@5 &MAP &NDCG@5 \\
\midrule
&\textbf{NA} &52.0 &25.0 &33.8 &43.3 &69.4 &66.2 &72.9 &15.9 &29.5 &20.6 &21.3 &32.6 &31.5 &35.8 \\
\midrule
\midrule
\multirow{2}{*}{IT} &\textbf{A} &51.6 &24.8 &33.5 &43.0 &71.1 &67.6 &74.2 &16.0 &29.7 &20.8 &21.5 &32.7 &32.0 &36.2 
\\
&\textbf{B} &51.4 &24.7 &33.3 &42.7 &71.0 &67.4 &74.0 &15.9 &29.5 &20.6 &21.4 &32.8 &32.0 &36.0
\\
&\textbf{C} &51.8 &24.9 &33.6 &43.5 &70.1 &67.9 &74.1 &15.7 &29.1 &20.4 &21.1 &32.0 &31.3 &35.8
\\
\midrule
\midrule
\multirow{2}{*}{ML} &\textbf{D} &42.8 &20.6 &27.8 &30.8 &58.1 &56.0 &62.7 &14.5 &26.9 &18.8 &19.6 &28.7 &27.8 &31.9 
\\
&\textbf{E} &42.6 &20.5 &27.7 &29.7 &60.0 &56.8 &62.7 &15.1 &28.1 &19.6 &20.5 &29.5 &29.0 &33.4
\\
\midrule
\midrule
\multirow{2}{*}{IR} &\textbf{F} &51.4 &24.7 &33.4 &42.9 &69.3 &66.5 &72.9 &15.6 &29.0 &20.3 &21.1 &32.4 &31.4 &35.8 
\\
&\textbf{G} &51.6 &24.8 &33.5 &42.6 &69.9 &67.6 &73.7 &15.2 &28.2 &19.7 &20.5 &32.1 &31.2 &35.3
\\
\bottomrule
\end{tabular}}
\end{table}

As shown in Table~\ref{tab:effect_prompts}, seven different prompt templates are selected to evaluate the effectiveness of prompts, which can be classified into three categories: instructive (correct legal prompts), misleading (wrongful legal prompts) and irrelevant (correct non-legal prompts). The performances of the experiments indicate that instructive prompts can improve performance by giving correct and informative indications of the global view of legal context. On the contrary, misleading prompts negatively impact the case retrieval accuracy. Compared with the other categories of prompts, irrelevant prompts slightly hurts the performance by adding irrelevant noisy information to the input. 

\begin{figure}[!t]
\centering
    \subfigure[SAILER]{
    \includegraphics[width=0.3\linewidth]{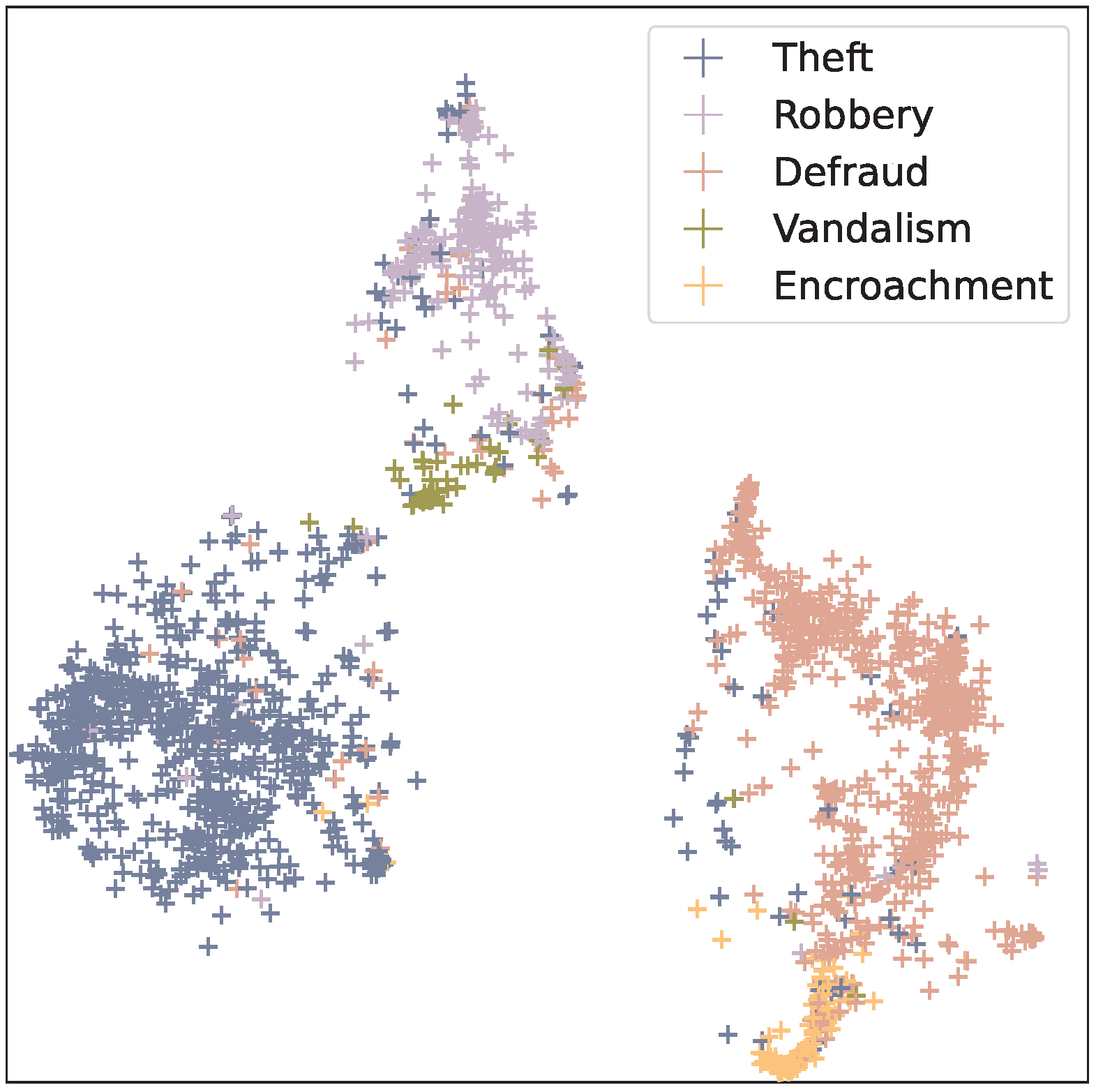}
    \label{fig:sailer_tsne}
    }
    \subfigure[SAILER + PromptCase]{
    \includegraphics[width=0.3\linewidth]{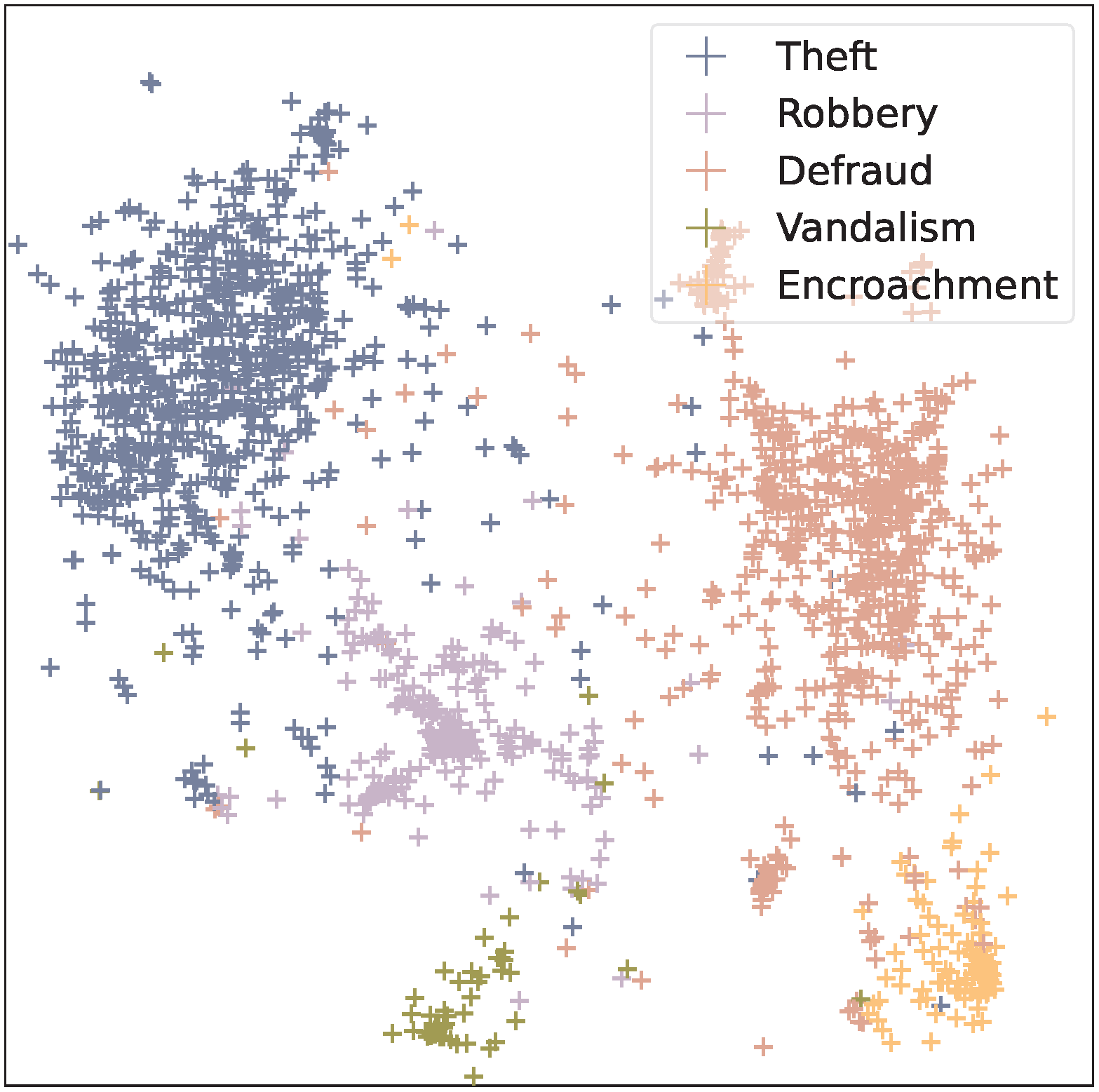}
    \label{fig:promptcase_tsne}
    }
\caption{Visulisation of case encodings with and without PromptCase for LeCaRD.}
\label{fig:tsne}
\end{figure}

\subsection{Visualisation Analysis}
To further prove the effectiveness of PromptCase input reformulation method, t-SNE~\cite{tsne} is used to visualise cases embeddings with and without PromptCase. Cases from five legally similar and difficult to distinguish charges of LeCaRD are selected to visualise in Fig.~\ref{fig:tsne}, including \textit{theft}, \textit{robbery}, \textit{defraud}, \textit{vandalism}, and \textit{encroachment}. All selected case embeddings are generated by the zero-shot SAILER model. As shown in Fig.~\ref{fig:sailer_tsne}, case embeddings generated by SAILER are classified into three clusters. Moreover, vandalism cases are wrongfully classified as robbery cases and encroachment cases are wrongfully classified as defraud cases. Compared with SAILER, adding PromptCase (as shown in Fig.~\ref{fig:promptcase_tsne}) makes cases embeddings evenly distributed as five clusters corresponding to five charges, which indicates the powerful discriminative ability and the ability to learn legal context information of PromptCase framework. 

\section{Conclusion}
This paper identifies the challenges in the existing LCR models about legal feature alignment and legal context preservation. To tackle these challenges, a novel legal case retrieval framework called PromptCase is introduced. In PromptCase, \textit{\textbf{legal facts}} and \textit{\textbf{legal issues}} are effectively extracted from the original case, which is further encoded with a prompt-based schema to generate an informative case representation. Extensive experiments are conducted on two benchmark datasets, which successfully demonstrate the superiority of PromptCase by achieving the best performance compared with state-of-the-art baselines.

\subsubsection{Acknowledgements} The work is supported by Australian Research Council CE200100025.

\bibliographystyle{splncs04}
\bibliography{mybibliography.bib}

\end{document}